\documentclass{article}
\usepackage{cite}
\usepackage{graphicx}
\usepackage{dcolumn}

\input{tcilatex}
\begin{document}

\date{}
\title{Comment on: ``Interaction of the magnetic quadrupole moment of a
non-relativistic particle with an electric field in a rotating frame. Ann.
Phys. 412 (2020) 168040''}
\author{Francisco M. Fern\'{a}ndez\thanks{%
fernande@quimica.unlp.edu.ar} \\
INIFTA, DQT, Sucursal 4, C.C 16, \\
1900 La Plata, Argentina}
\maketitle

\begin{abstract}
We analyze a recent treatment of the interaction of a magnetic quadrupole
moment with a radial electric field for a non-relativistic particle in a
rotating frame and show that the derivation of the equations in the paper is
anything but rigorous. The authors presented eigenvalues and eigenfunctions
for two sets of quantum numbers as if they belonged to the same physical
problem when they are solutions for two different models. In addition to it,
the authors failed to comment on the possibility of multiple solutions for
every set of quantum numbers.
\end{abstract}

In a recent paper\cite{HMM20} the authors studied the interaction of a
magnetic quadrupole moment with a radial electric field for a
non-relativistic particle in a rotating frame. They solved the
Schr\"{o}dinger equation for a model potential by means of a power-series
method and obtained the lowest eigenvalues and eigenfunctions. In this
Comment we analyze the derivation of the main equations and discuss their
solutions. We will not consider the validity of the model or its physical
utility, for this reason we skip most of the steps followed by the authors
to derive their main equations and outline just those that are relevant for
the discussion of the results.

The authors stated that ``In the geometric approach, the medium with a
disclination has the line element in cylindrical coordinates (in units such
that $c=1$), given by
\begin{equation}
ds^{2}=-dt^{2}+d\rho ^{2}+\alpha ^{2}\rho ^{2}d\varphi ^{2}+dz^{2},
\label{eq:metric}
\end{equation}
where $-\infty <z<\infty $, $-\infty <t<\infty $, $\rho \geq 0$ and $0\leq
\varphi \leq 2\pi $''. In this way they were able to derive a differential
operator that they called $\vec{\pi}$ that we do not show here because it is
not relevant for present purposes. In order to simplify the differential
equation for $\vec{\pi}^{2}$ the authors stated that ``If we consider $%
\partial _{\varphi }=i\ell $, $\partial _{z}=ik$ ...'' It is not clear if
the authors simply believe that those derivative operators are just
imaginary numbers and, besides, they did not indicate the possible values of
$\ell $ and $k$ (at this point). Notice that they also failed to say that $%
\hbar =1$ in the chosen units.

The authors chose the ``static scalar potential''
\begin{equation}
V(\rho )=a_{1}\rho +a_{2}\rho ^{2}-\frac{a_{3}}{\rho }+\frac{a_{4}}{\rho ^{2}%
},  \label{eq:potential}
\end{equation}
and later stated that ``The interaction is time-independent so that one can
write $\Psi (t,r,\varphi ,z)=e^{-i\left( \varepsilon t-\ell \varphi
-kz\right) }\psi (\rho )$.'' It is worth pointing out that the differential
equation for $\vec{\pi}^{2}$ does not contain a time derivative so that the
factor $e^{-i\varepsilon t}$ is unnecessary and the other two factors $%
e^{i\left( \ell \varphi +kz\right) }$ are also unnecessary because the
authors had already removed the derivatives $\partial _{\varphi }$ and $%
\partial _{z}$ in the unorthodox way indicated above.

In this way, and by means of a suitable change of variables, the authors
arrived at the eigenvalue equation
\begin{eqnarray}
&&\left[ \frac{d^{2}}{dr^{2}}+\frac{1}{r}\frac{d}{dr}-\frac{V_{-2}}{r^{2}}+%
\frac{V_{-1}}{r}-V_{1}r-r^{2}+W\right] \psi (r)=0,  \nonumber \\
&&V_{-2}=\frac{\ell ^{2}}{\alpha ^{2}}+2ma_{4},\;V_{-1}=\frac{2ma_{3}}{\sqrt{%
\eta }},\;V_{1}=\frac{2ma_{1}}{\eta ^{3/2}},\;W=\frac{\kappa ^{2}}{\eta },
\label{eq:eigen_eq}
\end{eqnarray}
where the parameters $a_{i}$, $\eta $ and $\kappa $ are given in the
authors' paper and will not be shown here. In order to solve this equation
the authors proposed the transformation
\begin{eqnarray}
\psi (r) &=&e^{-\frac{r^{2}}{2}-\frac{Cr}{2}}r^{\frac{\ell ^{2}}{\alpha ^{2}}%
+D}h(r),  \nonumber \\
C &=&\frac{2ma_{1}}{\eta ^{3/2}},\;D=-\frac{\ell ^{2}}{\alpha ^{2}}\pm
\alpha \sqrt{\ell ^{2}+2ma_{4}\alpha ^{2}},  \label{eq:authors_sol}
\end{eqnarray}
and stated that ``the positive sign is physically acceptable''. One can
easily verify that the correct behaviour at origin is $r^{s}$, where
\begin{equation}
s=\sqrt{V_{-2}}=\frac{1}{|\alpha |}\sqrt{\ell ^{2}+2ma_{4}\alpha ^{2}}\neq
\frac{\ell ^{2}}{\alpha ^{2}}+D.  \label{eq:s=}
\end{equation}
This \textit{misprint} is carried out throughout the paper.

The authors rewrote the resulting equation for $h(r)$
\begin{eqnarray}
&&h^{\prime \prime }(r)+\left( \frac{2\sqrt{V_{-2}}+1}{r}-2r\right)
h^{\prime }(r)-V_{1}h^{\prime }(r)+\frac{2V_{-1}-V_{1}\left( 2\sqrt{V_{-2}}%
+1\right) }{2r}h(r)+  \nonumber \\
&&\left( \frac{V_{1}^{2}}{4}-2\sqrt{V_{-2}}-2+W\right) h(r)=0,
\label{eq:diff_eq_h}
\end{eqnarray}
as the biconfluent Heun equation
\begin{equation}
H^{\prime \prime }(s)+\left( \frac{1+a}{s}-2s-b\right) H^{\prime }(s)+\left(
c-2-a-\frac{b|a+1|+d}{2s}\right) H(s)=0,  \label{eq:Heun}
\end{equation}
where the parameters $a$, $b$, $c$ and $d$ are given in the authors' paper
and here we only show the correct expression for $a$
\begin{equation}
a=2\sqrt{V_{-2}}=\frac{2}{|\alpha |}\sqrt{\ell ^{2}+2ma_{4}\alpha ^{2}}.
\label{eq:a=}
\end{equation}
Since $a>0$ then $|a+1|=a+1$ that greatly facilitates the calculation (it
seems that the authors did not realize this fact).

In order to solve the Heun equation the authors tried the power-series
\begin{equation}
H(s)=\sum_{n=0}c_{n}s^{n+p},  \label{eq:H(s)_series}
\end{equation}
and concluded that ``from the coefficient of $s^{p-2}$, we see that $p=0$ or
$p=-a$''. This analysis is unnecessary after having discussed the behaviour
of $h(r)$ at origin from which it follows that the physically acceptable
solution is in fact $p=0$. However, the authors commented on this point:
``For the sake of this paper, we shall consider only the solutions $p=0$
from Eq. (27)''. It seems that the authors believe that the other solution $%
p=-a$, already discarded previously, is suitable. The coefficients $c_{j}$
satisfy the three-term recurrence relation
\begin{eqnarray}
c_{j+2} &=&A_{j}c_{j+1}+B_{j}c_{j}=0,\;j=-1,0,1,\ldots ,\;c_{-1}=0,
\nonumber \\
A_{j} &=&\frac{2b(j+1)+b(a+1)+d}{2(j+2)(j+2+a)},\;B_{j}=\frac{a-c+2j+2}{%
(j+2)(j+2+a)}.  \label{eq:rec_rel}
\end{eqnarray}
In order to obtain polynomial solutions the authors chose the conditions $%
c-a=2\left( n_{0}+1\right) $ and $c_{n_{0}+1}=0$. From the former they
obtained an expression for the energy $\mathcal{E}_{n_{0},\ell }$ and the
latter tells us that not all the model parameters are independent. For
example, the authors decided to obtain $a$ in terms of the other parameters,
$n_{0}$ and $\ell $; that is to say $a_{n_{0},\ell }$. In this way the
authors stated that ``$a_{4}$ should be considered as $a_{4_{n_{0},\ell }}$%
''. Consequently, the potential (\ref{eq:potential}) should be written as $%
V_{n_{_{0}},\ell }(\rho )$ because it changes with the quantum numbers
through $a_{4}$. As a result each pair $\mathcal{E}_{n_{0},\ell }$, $\psi
_{n_{0},\ell }(\rho )$ obtained from the authors' procedure corresponds to
some model potential $V_{n_{_{0}},\ell }(\rho )$. Such quantum-mechanical
models are known as quasi-exactly solvable or conditionally solvable and
some variants of this model, even more general ones, have already been
treated before in a much more rigorous way\cite{D88, BCD17}. However, the
authors presented their explicit results $\mathcal{E}_{1,\ell }$, $\psi
_{1,\ell }(\rho )$ and $\mathcal{E}_{2,\ell }$, $\psi _{2,\ell }(\rho )$ as
if they were energies and states of the same model and as if the problem was
exactly solvable.

The fact that the model potential depends on the quantum numbers when we
force that kind of truncation condition is not the only feature of the
approach that they failed to mention. Another important point is that the
condition $c_{n_{0}+1}=0$ is a nonlinear equation that may have more than
one solution. In order to illustrate this point we substitute $2\left(
n_{0}+1\right) +a$ for $c$ in $c_{n_{0}+1}=0$ and solve for $a$. For
example, when $n_{0}=1$ we obtain
\begin{equation}
a_{1,\ell }^{\pm }=\frac{4-2b^{2}-bd\pm \sqrt{b^{4}-8b^{2}-8bd+16}}{b^{2}}.
\label{eq:a_1,l}
\end{equation}
Of course, we should choose a real, positive root. When $n_{0}=2$, $a$ is a
root of a cubic polynomial
\begin{eqnarray}
&&a^{3}b^{3}+a^{2}b\left( 9b^{2}+3bd-32\right) +a\left(
23b^{3}+18b^{2}d+3b\left( d^{2}-48\right) -32d\right) +  \nonumber \\
&&+15b^{3}+23b^{2}d+b\left( 9d^{2}-112\right) +d\left( d^{2}-48\right) =0.
\label{eq:cubic_a}
\end{eqnarray}
If, for a given $n_{0}$ there are more than one real positive root $%
a_{n_{0},\ell }$ then we would have eigenvalues and eigenfunctions for more
than one potential $V_{n_{0},\ell }$ for such pair of quantum numbers.

The problem of multiple solutions emerging from a truncation condition was
also overlooked by Bakke\cite{B14} in his calculation of bound states for a
Coulomb-type potential induced by the interaction between a moving electric
quadrupole moment and a magnetic field.

Summarizing: we have clearly seen that the derivation of the equations in
the paper by Hassanabadi et al\cite{HMM20} is anything but rigorous. They
presented eigenvalues and eigenfunctions for two sets of quantum numbers as
if they belonged to the same physical problem when they are solutions for
two different models. In addition to it, the authors failed to realize the
possibility of multiple solutions for every set of quantum numbers.

\section*{Addendum}

In what follows we analyze the reply to present Comment. With respect to our
criticism about the wrong behaviour at origin the authors stated that ``It
is a typo in one part of the article only and has no effect on the
results''. However, it is worth noticing that their wavefunctions (39) and
(43) already exhibit the wrong behaviour at origin (and we suppose that they
are part of their results). The authors appear to believe that the
expressions $\partial _{\varphi }=i\ell $ and $\partial _{z}=ik$ are correct.

The authors stated that they verified the correctness of their results with
other methods, such as ``Quasi-Exactly-Solvable method and Ansatz method'';
unfortunately, they did not give any reference and we do not know what they
exactly mean by such names. They seem to be something different from the
Frobenius method.

The main point is that the results given by the Frobenius method, followed
by a suitable truncation of the series, are not wrong by themselves. What is
wrong is the interpretation of such results. In the Comment we said (or, at
least, tried to) that the authors obtained eigenvalues for different model
potentials and presented them as if they were the spectrum of a single
problem. In this Addendum we expand on this issue and show revealing results.

For present discussion we rewrite the eigenvalue equation as
\begin{equation}
u^{\prime \prime }(x)+\frac{1}{x}u(x)-\frac{\gamma ^{2}}{x^{2}}u(x)-\frac{a}{%
x}u(x)-bxu(x)-x^{2}u(x)+Wu(x)=0,  \label{eq:eig_eq}
\end{equation}
where $\gamma $, $a$ and $b$ are real model parameters that have nothing to
do with the parameters in the equations discussed in the paper and Comment.
Only the form of the equation is the same. This eigenvalue equation has
square integrable solutions
\begin{equation}
\int_{0}^{\infty }\left| u(x)\right| ^{2}x\,dx<\infty ,
\label{eq:bound_states}
\end{equation}
for all $-\infty <a,b<\infty $ for an infinite number of discrete values of $%
W(a,b)$. Such eigenvalues satisfy the Hellmann-Feynman theorem\cite{F39}
\begin{equation}
\frac{\partial W}{\partial a}=\left\langle \frac{1}{x}\right\rangle >0,\;%
\frac{\partial W}{\partial b}=\left\langle x\right\rangle >0.  \label{eq:HFT}
\end{equation}

In what follows we try to solve the eigenvalue equation (\ref{eq:eig_eq}) by
means of the Frobenius method and the ansatz
\begin{equation}
u(x)=x^{s}\exp \left( -\frac{b}{2}x-\frac{x^{2}}{2}\right)
P(x),\;P(x)=\sum_{j=0}^{\infty }c_{j}x^{j},\;s=\left| \gamma \right| .
\label{eq:ansatz}
\end{equation}
The expansion coefficients $c_{j}$ satisfy the three-term recurrence
relation
\begin{eqnarray}
c_{j+2} &=&A_{j}c_{j+1}+B_{j}c_{j},\;j=-1,0,1,2,\ldots ,\;c_{-1}=0,\;c_{0}=1,
\nonumber \\
A_{j} &=&\frac{2a+b\left( 2j+2s+3\right) }{2\left( j+2\right) \left[
j+2\left( s+1\right) \right] },\;B_{j}=\frac{4\left( 2j+2s-W+2\right) -b^{2}%
}{4\left( j+2\right) \left[ j+2\left( s+1\right) \right] }.  \label{eq:TTRR}
\end{eqnarray}
If the truncation condition $c_{n+1}=c_{n+2}=0$, $c_{n}\neq 0$, $%
n=0,1,\ldots $, has physically acceptable solutions for $a$, $b$ and $W$
then we obtain exact eigenfunctions because $c_{j}=0$ for all $j>n$. This
truncation condition is equivalent to $B_{n}=0$, $c_{n+1}=0$ or
\begin{equation}
W_{s}^{(n)}=2\left( n+s+1\right) -\frac{b^{2}}{4},\;c_{n+1}(a,b)=0,
\label{eq:trunc_cond}
\end{equation}
where the second condition determines a relationship between the parameters $%
a$ and $b$. On setting $W=W_{s}^{(n)}$ the coefficient $B_{j}$ takes a
simpler form:
\begin{equation}
B_{j}=\frac{2\left( j-n\right) }{\left( j+2\right) \left[ j+2\left(
s+1\right) \right] }.
\end{equation}
Notice that the truncation condition does not provide all the solutions but
only those for which the parameters $a$ and $b$ exhibit certain relations.
The reason is that this problem is not exactly solvable, as the authors
appear to believe, but quasi-exactly solvable or conditionally solvable (see
\cite{CDW00,AF20,F20b,F20c} and, in particular, the remarkable review \cite
{T16} and references therein for more details).

As an illustrative example we consider the eigenvalue equation (\ref
{eq:eig_eq}) with $b=1$. In this case $c_{n+1}(a,1)=0$ is a polynomial
function of $a$ of degree $n+1$ and it can be proved that all the roots $%
a_{s}^{(n,i)}$, $i=1,2,\ldots ,n+1$, are real\cite{CDW00,AF20}. For
convenience we arrange the roots so that $a_{s}^{(n,i)}>a_{s}^{(n,i+1)}$ and
stress the point that all of them correspond to the same eigenvalue $%
W_{s}^{(n,i)}=W_{s}^{(n)}$. It is important to realize that the eigenvalue $%
W_{s}^{(n)}$ is common to a set of different quantum-mechanical problems
because the potential depends on $a$. The origin of the authors'
misconception can be traced back to this obvious fact. For example, the
eigenvalues $\mathcal{E}_{n_{0},l}$ obtained by them correspond to different
quantum-mechanical problems and are, consequently, meaningless. The
polynomial solutions
\begin{equation}
u_{s}^{(n,i)}(x)=x^{s}\exp \left( -\frac{x^{2}}{2}\right)
P_{s}^{(n,i)}(x),\;P_{s}^{(n,i)}(x)=\sum_{j=0}^{n}c_{j,s}^{(n,i)}x^{j},\;s=%
\left| \gamma \right| ,  \label{eq:poly_sol_CH}
\end{equation}
share the same eigenvalue $W_{s}^{(n)}$ and also correspond to different
quantum-mechanical problems.

The actual eigenvalues $W_{\nu ,s}(a)$, $\nu =0,1,\ldots $, $W_{\nu
,s}<W_{\nu +1,s}$, of equation (\ref{eq:eig_eq}) (for a given value of $b$)
are curves in the $a-W$ plane. It follows from the Hellmann-Feynman theorem (%
\ref{eq:HFT}) that $\left( a_{s}^{(n,i)},W_{s}^{(n)}\right) $ is a point on
the curve $W_{i-i,s}(a)$. In order to verify this fact we need the actual
eigenvalues $W_{\nu ,s}(a)$ that we have to obtain by means of a suitable
approximate method because the eigenvalue equation (\ref{eq:eig_eq}) is not
exactly solvable\cite{AF20,T16}. Here, we resort to the well known
Rayleigh-Ritz variational method that is known to yield upper bounds to all
the eigenvalues\cite{P68} and, for simplicity, choose the non-orthogonal
basis set of Gaussian functions $\left\{ \varphi _{j,s}(x)=x^{s+j}\exp
\left( -\frac{x^{2}}{2}\right) ,\;j=0,1,\ldots \right\} $.

\begin{figure}[tbp]
\begin{center}
\includegraphics[width=9cm]{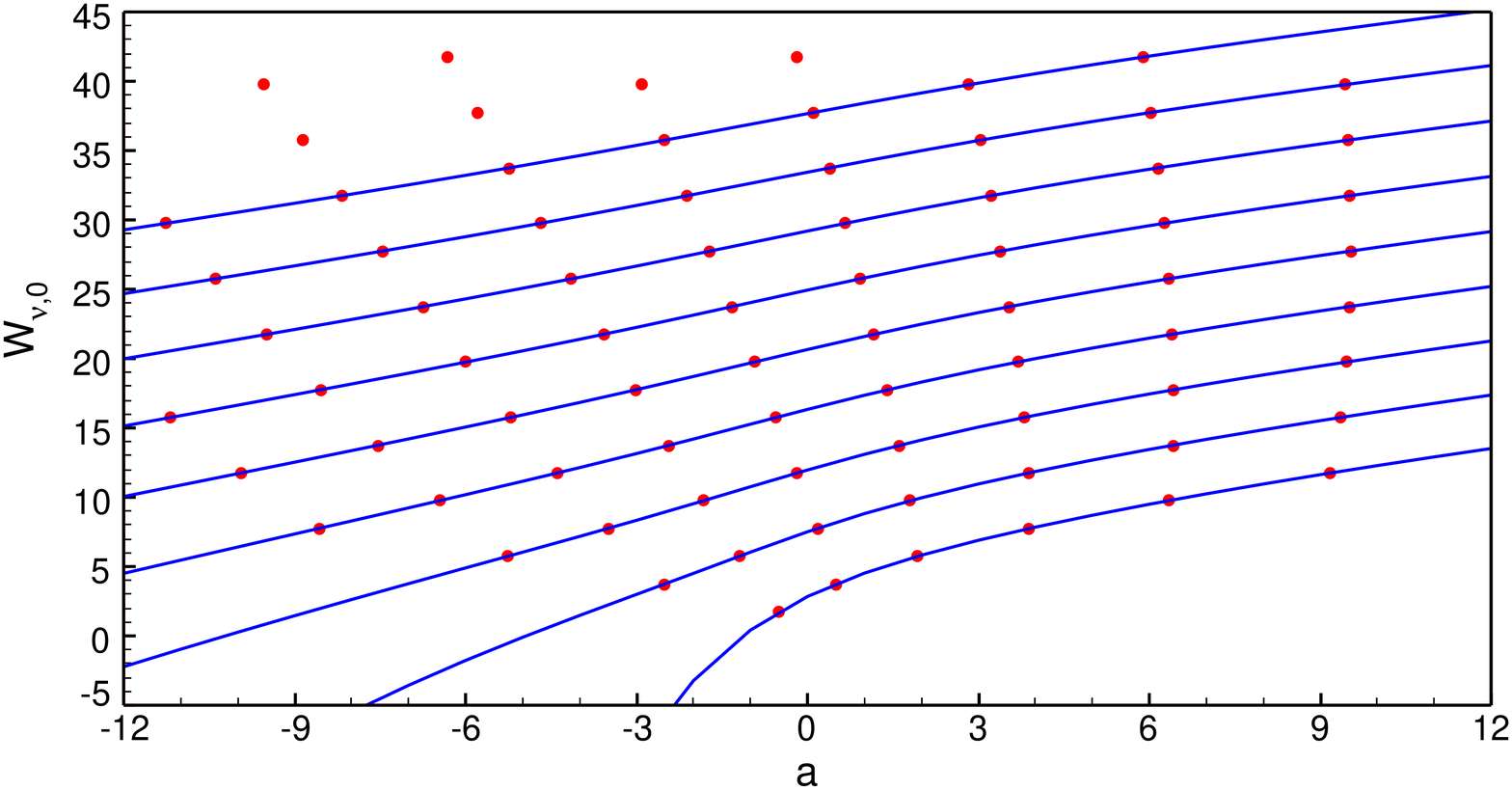}
\end{center}
\caption{Eigenvalues $W_0^{(n)}(a,1)$ from the truncation condition (red
points) and $W_{\nu,0}(a)$ obtained by means of the variational method (blue
lines)}
\label{Fig:Wb1g0}
\end{figure}

In order to facilitate the variational calculations we choose $s=0$ in what
follows. Figure~\ref{Fig:Wb1g0} shows several eigenvalues $W_{0}^{(n)}$
given by the truncation condition (red points) and the lowest actual
eigenvalues $W_{\nu ,0}(a)$ obtained from the variational method (blue
lines). We see that there are solutions to the eigenvalue equation (\ref
{eq:eig_eq}) for all values of $a$, that each $W_{\nu ,0}(a)$ is a
continuous function of $a$ that satisfies the Hellmann-Feynman theorem (\ref
{eq:HFT}) and that each pair $\left( a_{0}^{(n,i)},W_{0}^{(n)}\right) $ is a
point on those curves as argued above. Any vertical line starting from a
given value of $a$ will pass through no more that one red point. It means
that the truncation condition yields only one eigenvalue and just for a
particular model potential. We realize that the eigenvalues obtained by
Hassanabadi et al\cite{HMM20} have no physical meaning unless one connects
the points $\left( a_{s}^{(n,i)},W_{s}^{(n)}\right) $ properly.

\end{document}